\documentclass{iaus}
\usepackage{graphicx}

\title[Chemical evolution of Seyferts] 
{Chemical evolution of Seyfert~galaxies}

\author[S.K. Ballero et al.]   
{Silvia K. Ballero$^{1,2}$, Francesca Matteucci$^{1,2}$, Luca
  Ciotti$^{3}$}

\affiliation{$^1$Dipartimento di Astronomia, Universit\`a di Trieste,
  via G.B. Tiepolo 11, 34124 Trieste, Italy \\[\affilskip]
  $^2$INAF, Osservatorio Astronomico di Trieste, via G.B. Tiepolo 11,
  34124 Trieste, Italy \\[\affilskip] 
  $^3$Dipartimento di Astronomia, Universit\`a di Bologna, via
  Ranzani 1, 40127, Bologna, Italy}

\pubyear{2007}
\volume{245}  
\pagerange{??--??}
\date{29/08/2007 and in revised form ??}
\jname{Proceedings: Formation and Evolution of Galaxy Bulges}
\editors{M. Bureau, E. Athanassoula \& B. Barbuy, eds.}
\begin{document}

\maketitle

\begin{abstract}
We computed the chemical evolution of Seyfert galaxies, residing in
spiral bulges, based on an updated model for the Milky Way bulge  
with updated calculations of the Galactic potential and of
the feedback from the central supermassive black hole (BH) in a
spherical approximation. 
We followed the evolution of bulges of masses $2\times
10^{9}-10^{11}M_{\odot}$ by scaling the star-formation efficiency and
the bulge scalelenght as in the inverse-wind scenario for
ellipticals. 
We successfully reproduced the observed relation between the 
BH mass and that of the host bulge, and the observed peak
nuclear bolometric luminosity. 
The observed metal overabundances are easily achieved, as well as the
constancy of chemical abundances with the redshift.    
\end{abstract}

\firstsection 
\section{The evolution model}

We based our analysis on the model by \cite{Ballero07}, which holds
for a bulge of $M_{b}=2\times 10^{10}M_{\odot}$. 
In order to analyse bulges of different masses, we chose to keep the
IMF and infall timescales constant and to scale the effective radius
and the star-formation efficiency following the inverse-wind scenario
of \cite{Matteucci94}; therefore, for bulges of $2\times 10^{9}$ and
$10^{11}M_{\odot}$ we adopted a star-formation efficiency of 11 and 50
Gyr$^{-1}$ and an effective radius of 1 and 4 kpc, respectively.

We define a displacement radius $r_t$ where the gas contained at
$r<r_t$ is carried by~the wind, and we adopt $r_t = 3R_e$.
The gas binding energy $\Delta E_b$ is made up by the contribution
of three components, a \cite{Hernquist90} distribution for the bulge
stellar component, an isothermal dark matter halo with circular
velocity $v_c$ and a razor-thin exponential disk.
These components are represented by the terms at the r.h.s. of the
equation:
\begin{equation}
\Delta E_b = \frac{GM_gM_b}{r_b}\times \Delta \tilde{f}_b (\delta)
       + M_gv_c^2\times \Delta \tilde{f}_{DM} (\delta)
       + \frac{GM_gM_d}{r_d} \times \Delta \tilde{f}_d (\delta,r_d)
\end{equation}
where $M_g$ is the bulge gas mass, $r_d$ is the disk scalelenght, $r_b =
R_e/1.8$ and $\delta=r_t/r_b$. 

The supernova feedback is calculated as in \cite{Pipino02}. 
We consider a spherically accreting BH, at a rate
$\dot{M}_{BH}$ given by the minimum between the Bondi and Eddington
rate (\cite{Sazonov05}). 
The bolometric luminosity emitted by the BH is then
$L_{bol} = \eta c^2\dot{M}_{BH}$ ($\eta=0.1$, \cite{Yu02}) and we
assume that the energy released is a fraction $f=0.05$ 
(\cite{DiMatteo05}) of $L_{bol}$ integrated over the~timestep.

\section{Results}

We showed that the BH feedback is in most cases not significant
in triggering the galactic wind with respect to the supernova
feedback.
The predicted BH masses are in agreement with measurements of BH
masses inside Seyferts (e.g. \cite{Peterson03}) and reproduce fairly
well the relations of \cite{McLure02} and \cite{Marconi03}.
The predicted bolometric luminosities lie in the range
$10^{42}-10^{44}$ ergs~s$^{-1}$, in agreement with recent estimates
(e.g. \cite{Wang05}), although we could not reproduce a true
quiescence after the peak due to the lack of a hydrodynamical
treatment. 

Solar metallicity is reached in a very short time ($\lesssim 10^{8}$ yr). 
The huge metallicities inferred from observations (\cite{Fields05b})
are thus easily achieved, with more massive bulges giving 
rise to higher metallicities (up to $7$ times solar).
After the first $\sim 3\times10^8$ years, the bulge ISM reaches
overabundances of up to 10 times solar for N, Fe, Si, 5 times solar
for Mg and 3 times solar for C, O, roughly consistent with the
estimates for Fe, N and O in the broad and intrinsic narrow line
regions of Seyferts (e.g. \cite{Storchi96}; \cite{Ivanov03}).
The slightly supersolar [Fe/Mg] and its weak dependence on the bulge
luminosity (mass) are recovered (\cite{Dietrich03}). 
The mean value of the [N/C] ratio is very sensitive on the bulge mass,
due to its sensitivity on metallicity. 
The predicted values (N/C $\sim1.5-4$(N/C)$_{\odot}$) agree with the
estimates of \cite{Fields05a}.

\begin{figure}
\centering
  \includegraphics[width=.33\textwidth]{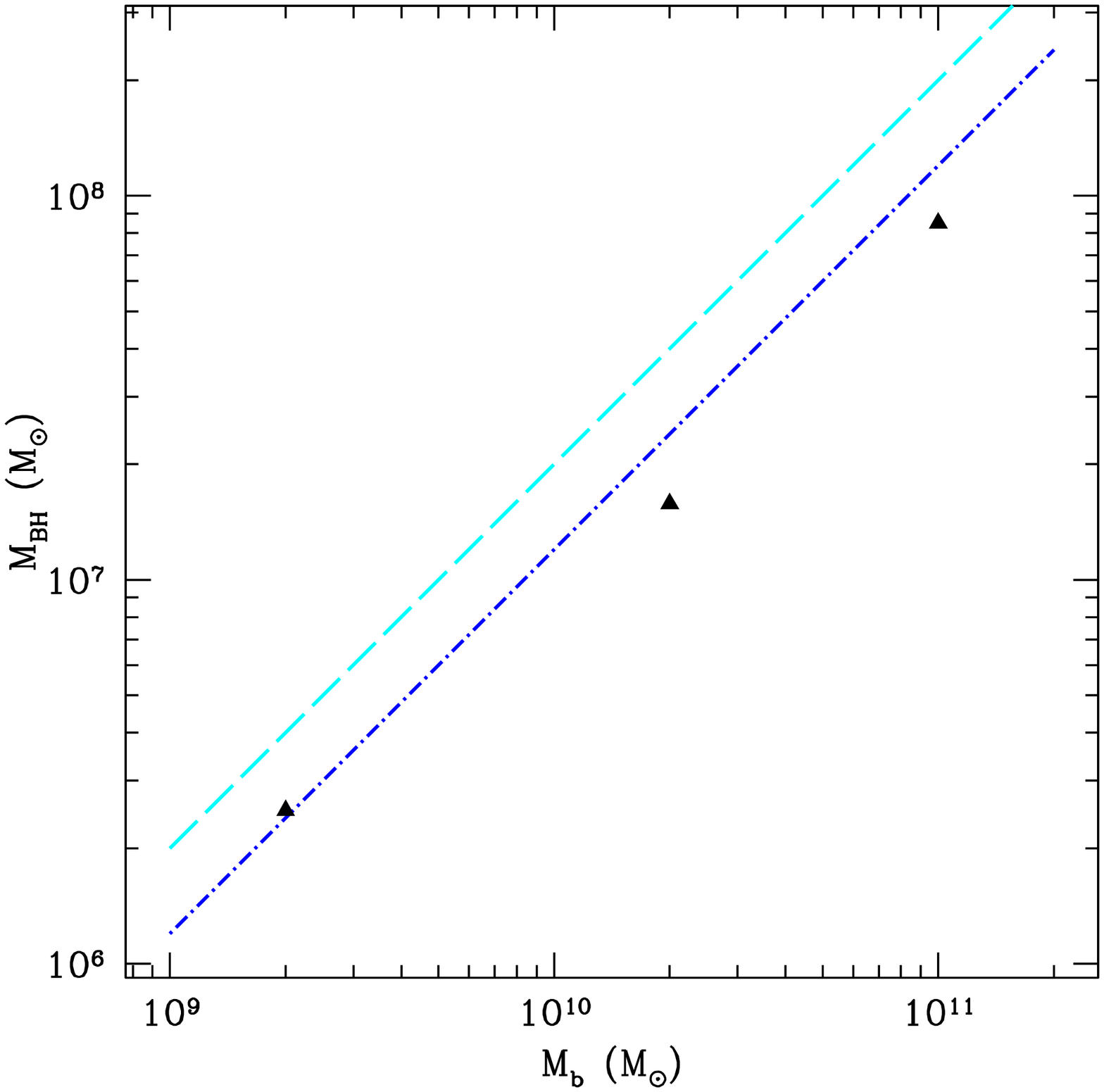}%
  \includegraphics[width=.33\textwidth]{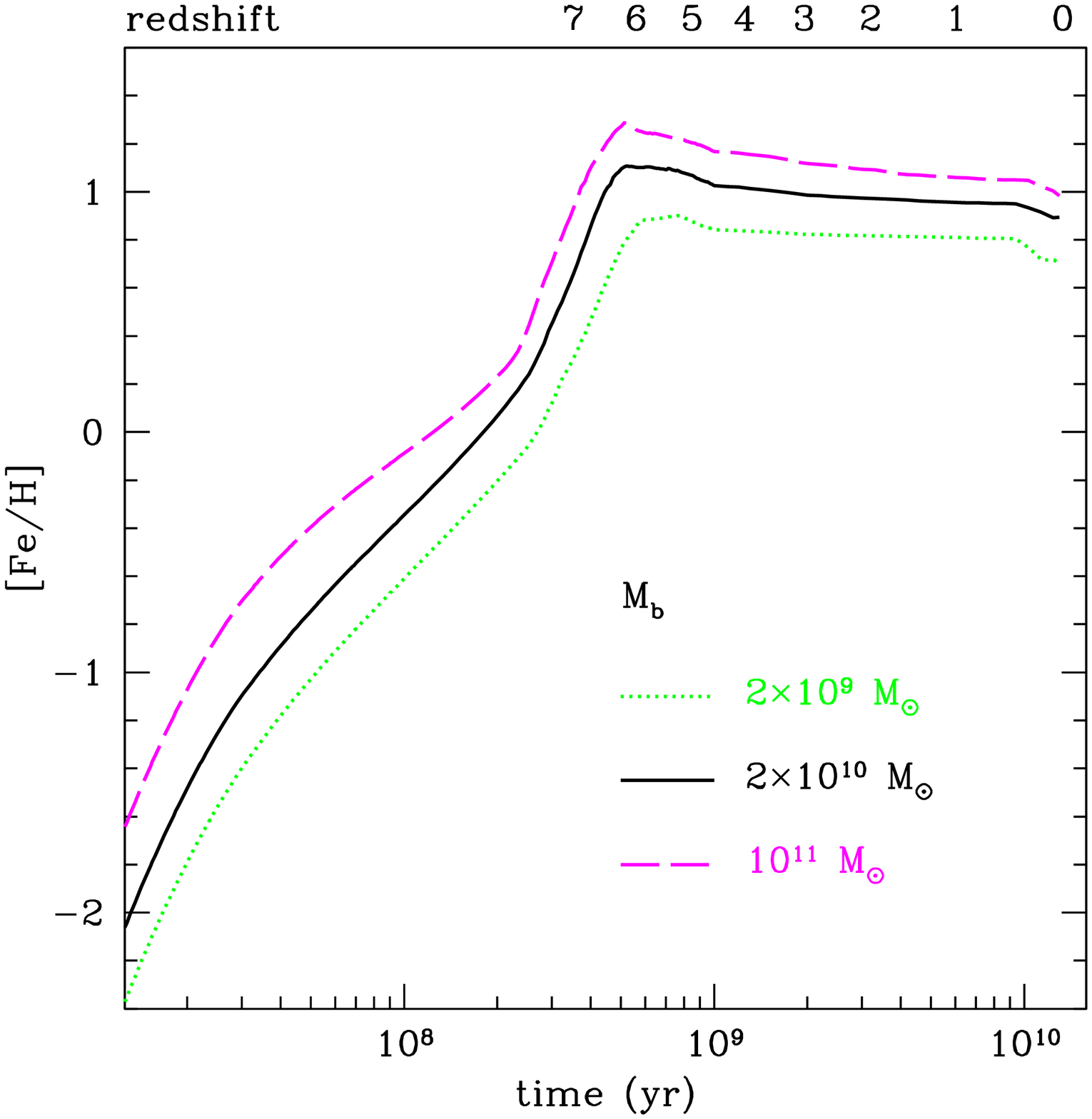}%
  \includegraphics[width=.33\textwidth]{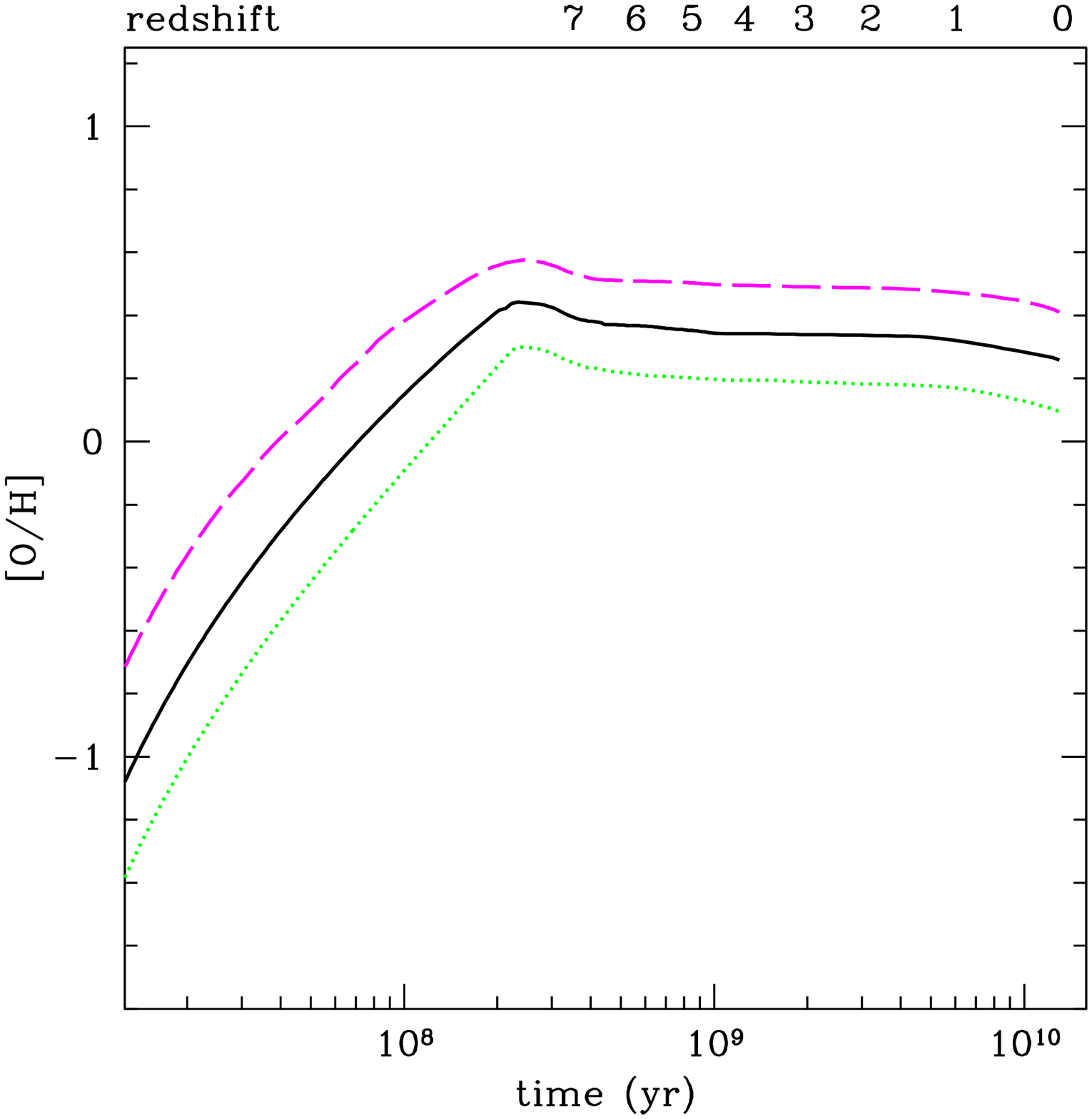}
  \caption{\textit{Left panel}: predicted $M_{BH}$ \textit{vs.} $M_b$
    relation compared with the observations of Marconi \& Hunt (2003,
    dashed-dotted line) and McLure \& Dunlop (2002, long-dashed line).  
    \textit{Middle and right panel}: Evolution with time and redshift
    $z$ of the [Fe/H] and [O/H] abundance ratios; $z$ is derived for a
    $\Lambda$CDM cosmology with $H_{0} = 65$ km s$^{-1}$ Mpc$^{-1}$,
    $\Omega_{M} = 0.3$, $\Omega_{\Lambda} = 0.7$, and
    $z_{f}\simeq10$.} 
\end{figure}

\end{document}